\documentclass[useAMS,usenatbib]{mn2e}
\usepackage{graphicx}
\def\ltsima{$\; \buildrel < \over \sim \;$}
\def\simlt{\lower.5ex\hbox{\ltsima}}
\def\gtsima{$\; \buildrel > \over \sim \;$}
\def\simgt{\lower.5ex\hbox{\gtsima}}
\def\igr{IGR J17511--3057}
\def\xte{XTE J1751--305}
\def\saxj{SAX J1808.4--3658}
\def\rxte{\it RXTE}
\def\xmm{\it XMM-Newton}

\title[X-ray spectrum of the newly discovered AMSP {\igr}]{The X-ray spectrum of the newly discovered accreting millisecond pulsar {\igr}}
\author[A. Papitto et al.]{A.~Papitto$^{1},^{2}$\thanks{E-mail:
apapitto@oa-cagliari.inaf.it}, A.~Riggio$^{1}$, T.~Di Salvo$^{3}$, L.~Burderi$^{2}$, A.~D'A\`i$^{3}$,  R.~Iaria$^{3}$, E.~Bozzo$^{4}$,  \newauthor and M.~T. Menna$^{5}$\\ 
$^{1}$ INAF - Osservatorio Astronomico di Cagliari, Poggio dei Pini, Strada 54, 09012 Capoterra (CA), Italy \\
$^{2}$ Dipertimento di Fisica, Universit\'a degli Studi di Cagliari, SP Monserrato-Sestu, KM 0.7, 09042 Monserrato, Italy\\
$^{3}$ Dipartimento di Scienze Fisiche ed Astronomiche, Universit\'a di Palermo, via Archirafi 36, 90123 Palermo, Italy\\
$^{4}$ ISDC data centre for astrophysics, University of Geneva, chemin d´Ecogia, 16 1290 Versoix, Switzerland\\
$^{5}$ INAF Osservatorio Astronomico di Roma, via Frascati 33, 00040 Monteporzio Catone, Italy}

\begin{document}
                                   

\maketitle

\label{firstpage}

\begin{abstract}

We report on a 70ks {\xmm} ToO observation of the newly discovered
accreting millisecond pulsar, {\igr}. Pulsations at 244.8339512(1) Hz
are observed throughout the outburst with an RMS pulsed fraction of
14.4(3)$\%$. Pulsations have been used to derive a precise solution
for the P$_{orb}=12487.51(2)$s binary system.  The measured mass
function indicates a main sequence companion star with a mass between
0.15 and 0.44 M$_{\odot}$.

The {\xmm} 0.5--11 keV spectrum of {\igr} can be modelled by at least
three components, which we interpret, from the softest to the hardest,
as multicoloured disc emission, thermal emission from the neutron star
surface and thermal Comptonization emission. Spectral fit of the
{\xmm} data and of the {\it RXTE} data, taken in a simultaneous
temporal window, well constrain the Comptonization parameters: the
electron temperature, kT$_e$=51$^{+6}_{-4}$ keV, is rather high, while
the optical depth ($\tau$=1.34$^{+0.03}_{-0.06}$) is moderate.

The energy dependence of the pulsed fraction supports the
interpretation of the cooler thermal component as coming from the
accretion disc, and indicates that the Comptonizing plasma surrounds
the hot spots on the neutron star surface, which in turn provide the
seed photons.  Signatures of reflection, such as a broadened iron
K$\alpha$ emission line and a Compton hump at $\sim 30$ keV, are also
detected.  We derive from the smearing of the reflection component an
inner disc radius of $\simgt$ 40 km for a 1.4 M$_{\odot}$ neutron star, and an
inclination between 38$^{\circ}$ and 68$^{\circ}$.

{\xmm} also observed two type-I X-ray bursts, whose fluence and
recurrence time suggest that the bursts are ignited in a nearly pure
helium environment.  No photospheric radius expansion is observed,
thus leading to an upper limit on the distance to the source of 10
kpc. A lower limit of 6.5 kpc can be also set if it is assumed that
emission during the decaying part of the burst involves the whole
neutron star surface.  Pulsations are observed during the burst decay
with an amplitude similar to the persistent emission. They are also
compatible with being phase locked to pre-burst pulsations, suggesting
that the location on the neutron star surface where they are formed
does not change much during bursts.

\end{abstract}


\section{Introduction}

The discovery of 244.8 Hz pulsations in the X-ray emission of the
newly discovered transient source, {\igr} \citep{Mrk09a}, has brought
to  thirteen the number of Accreting Millisecond Pulsars (hereafter
AMSP) discovered so far.  Such high spin frequencies are believed to
ensue from a long phase ($\simgt 10^8$ yr) of accretion of mass and
angular momentum onto the neutron star (NS), according to the
so-called recycling scenario \citep[][]{BhtvdH91}.

In this paper we mainly focus on the spectral properties of this
source. The broadband continuum of AMSP has been invariably found to
be dominated by a power-law like hard emission with a cut off at an
energy between 30 and 60 keV .  This emission is interpreted as
Comptonization of soft photons in a hot plasma \citep[see][and
  references therein, for a review of the spectral properties of
  AMSP]{Pou06}.  At energies of $\sim$1 keV two soft components are
also generally found (\citealt{GrlPou05}, GP05 hereinafter,
\citealt{Pap09,Pat09}). The cooler one is attributed to the accretion
disc emission, while the hotter is interpreted as thermal emission of
the hotspots on the NS surface. In this scenario, the hotspot thermal
emission constitutes the bulk of soft photons that Compton
down-scatter the hot plasma.  The X-ray spectrum of {\igr} is found
consistent with this phenomenological description and the outlined
physical interpretation fits well the observed spectral properties.

\begin{figure}
 \includegraphics[angle=270,width=\columnwidth]{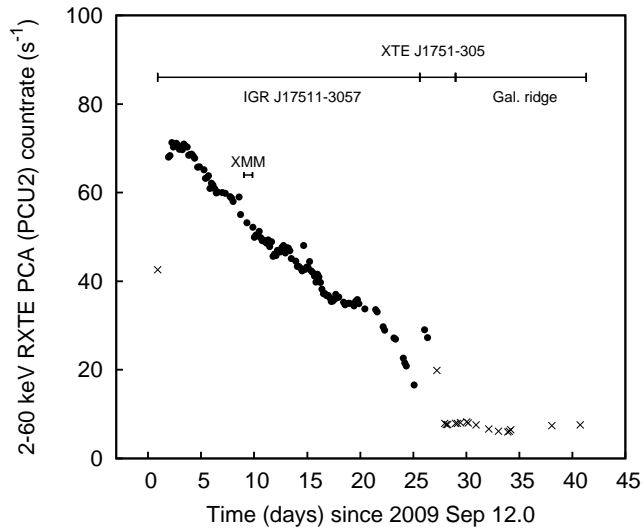}
 \caption{2--60 keV light curve of the {\igr} outburst as observed by
   the PCU2 of the PCA aboard {\rxte}. Circles and crosses refer to
   observations pointed towards {\igr} (Obsid P94041) and {\xte}
   (Obsid P94042), respectively. As the two PCA fields of view greatly
   overlap, the detected pulse frequency is used to discriminate from
   which source the emission observed by the PCA was actually coming
   from (see text for details). To which source the emission is
   attributed is indicated at the top of the figure. The time at which
   the {\xmm} observation was performed is also shown.}
\label{fig:lc}
\end{figure}

A couple of AMSP have also showed the typical clues of disc
reflection, a Compton hump at $\sim30$ keV \citep{GrlDonBar02} and a
K$\alpha$ emission iron line \citep{Pap09,Cac09,Pat09}. In particular
the line observed from {\saxj} is compatible with the typically
broadened shape caused by the relativistic Keplerian motion of the
inner rings of the accretion disc, immersed in the deep gravitational
well of the compact object \citep{Fab89}. The observation of similar
broadened lines from AMSPs has a peculiar importance in the context of
NS accretors.  It may constrain in fact where the magnetosphere breaks
off the accretion disc flow and lifts off matter to the accretion
columns.

The spectral and timing capabilities of {\xmm} already proved
fundamental to investigate the X-ray emission and pulse properties of
AMSPs especially at low energies, as well as to detect iron emission
lines at a high statistics. To these ends, we have obtained a 70ks
Target of Opportunity (ToO) observation of {\igr}, performed roughly
one week after the outburst onset. We report on the spin and orbital
properties of the source in Sec. \ref{sec:pulse}, and on its spectrum
in Sec.\ref{sec:xmmspectrum}.  We also show how it is possible to take
advantage of a simultaneous Rossi X-ray Timing Explorer ({\it RXTE})
observation, in order to constrain its high energy emission
(Sec. \ref{sec:rxtexmm}).

Also two type I X-ray bursts were caught by {\xmm}, making this
observation particularly rich (Sec.\ref{sec:burst}).  We observed
burst oscillations during the bursts, at the same frequency of the
non-burst emission.  It is of key importance to observe burst
oscillations from sources, the rotational state of which is known at a
great accuracy. The comparison with the properties of the non-burst
pulsations may in fact assess the physical mechanism that produces
oscillations during bursts.  These have already been observed to be
strongly phase locked to the non-burst oscillations, possibly
suggesting that the similarities in their formation are more profound
than expected \citep{Wts08}.

\section[]{Observations and data analysis}

The X-ray transient {\igr} has been discovered by {\it INTEGRAL} on
2009 September 21.140, during its Galactic Bulge monitoring
\citep{Bald09}. At that time also the {\it RXTE} Proportional Counter
Array (PCA) detected a rising X-ray flux activity from the Galactic
Bulge. This emission was tentatively attributed to two known nearby
transient sources, {\xte}, an AMSP spinning at 435 Hz \citep{Mark02},
and GRS 1747--312. {\rxte} therefore pointed at the position of {\xte}
to discover that the emission instead originated by a source spinning
at $\sim$245 Hz. This made {\igr} the twelfth AMSP discovered
\citep{Mrk09a}. A follow up {\it Swift} ToO observation constrained
the source position at a few arcsec accuracy, obtaining a value
$\sim$20 arcmin away from {\xte}, thus confirming the discovery of a
new source \citep{Bozzo09}. A {\it Chandra} observation further
refined the source position \citep{Now09}, giving the most accurate
available estimate as RA=17h~51'~08.66'', DEC=-~30$^{\circ}$~57'~
41.0'' (1 sigma error of 0.6 arcsec), which is the one we consider in
the following.

\subsection{\xmm}

\begin{figure}
 \includegraphics[angle=270.0,width=\columnwidth]{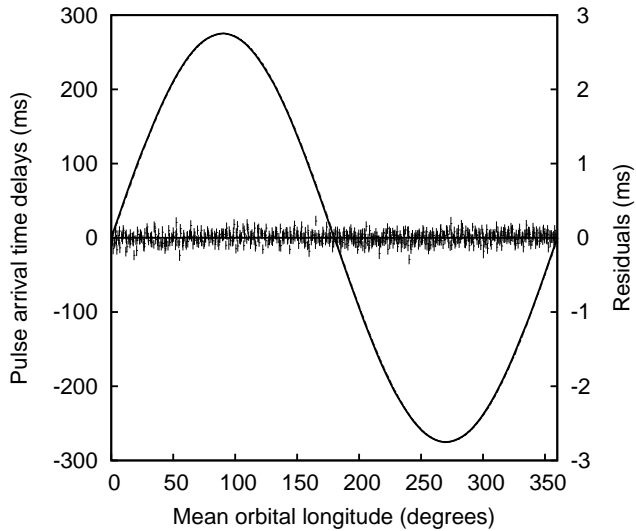}
 \caption{Pulse phase time delays induced by the orbital motion as a
   function of the mean orbital longitude. Longitude 0$^{\circ}$ refers to
   the passage of the pulsar at the ascending node of the
   orbit. Residuals with respect to the best fitting orbit model are
   also over-plotted at a magnified scale (rightmost y-axis). The
   reduced chi squared of the model is 1.15 over 684 degrees of
   freedom.  }
\label{fig:orbit}
\end{figure}

{\xmm} observed {\igr} as a ToO observation for 70ks starting on 2009
September 21.037, $\sim$ 9 days after the source discovery. The {\xmm}
observation is indicated by a horizontal bar in Fig.\ref{fig:lc},
where the {\it RXTE} PCA light curve of the outburst is plotted.
 
The EPIC-pn camera operated in timing mode to allow the temporal
resolution (30 $\mu$s) needed to study the millisecond variability of
the source, and to avoid spectral deformations due to pile-up. Also
EPIC-MOS2 camera observed in timing mode, while EPIC-MOS1 operated in
Small Window to provide an image of the source.  No evident background
due to soft proton flares was recorded during the observation, and the
full exposure window is retained for scientific purposes. {\xmm}
detected two type I X-ray bursts during its pointing.  When analysing
the {\it persistent} emission\footnote{Even if the source is an X-ray
  transient, we refer to {\it persistent} emission as the main body of
  the outburst emission, and to {\it burst} emission when the source
  exhibits thermonuclear flashes, so called type I X-ray bursts.}, we
discarded 20s prior and 110s after the burst onset. This choice is
fairly conservative as the e-folding factor of the burst decay is
$\tau\simeq11$s (see Sec.\ref{sec:burst}).

{\xmm} data have been extracted and reduced using SAS v.9.0.  A
concatenated and calibrated event file has been created with the task
epproc, specifying the Chandra coordinates of the source.  The EPIC-pn
spectrum has been extracted from a 13 pixels wide region around the
source position (RAWX=36), equivalent to 53.3 arcsec (which should
encircle more than 90 per cent of the source counts up to 9
keV\footnote{See {\xmm} Users handbook, issue 2.6, available at
  http:// xmm.esac.esa.int/external/xmm\_user\_support.})  A similar
stripe centred on RAWX=10 has been used to estimate the background.
The background subtracted 1.4--11 keV\footnote{see
  Sec.\ref{sec:xmmspectrum} for details about the energy interval
  chosen for the purposes of a spectral analysis.} {\it persistent}
count rate recorded by the EPIC-pn is 43.9 c/s on the average.  Pile
up is therefore not an issue for EPIC-pn data.  Spectral channels have
been re-binned to have at least three channels per resolution element
and 25 counts per channel.

EPIC-MOS1 data have been extracted using the task emproc to provide a
101''$\times$101'' image of the source. Using the task edetect\_chain,
we estimate the source position as RA=17h 51m 08.55s,
DEC=-30$^{\circ}$ 57' 41.7'', with an uncertainty of 4.3 arcsec.  This
is consistent with the more accurate Chandra determination. EPIC-MOS 2
data have a significantly lower statistics than that of EPIC-pn and
are therefore discarded from the spectral analysis.

Reflection Grating Spectrometers (RGS) operated in standard
spectroscopy mode and data have been reduced using the task
rgsproc. We consider only first order spectra, re-binned in order to
have at least 25 counts per channel. The average {\it persistent}
count rate is 0.4 and 0.5 c/s for RGS1 and RGS2, respectively.

All spectral fits were performed using XSPEC v.12.5.1

\subsection{\rxte}
\label{sec:rxte}

Similarly to the other AMSPs, only one third of the X-ray flux of
{\igr} is emitted in the energy range covered by {\xmm}. In order to
better constrain the broadband X-ray emission, we take advantage of
two {\it RXTE} observations performed simultaneously to the {\xmm}
pointing, and amounting to an exposure of 8.8ks (Obs. 94041-01-02-08
and Obs. 94041-01-02-18).

{\rxte} observations have also been used to extract a light curve of
the whole outburst. Starting on 2009 September 13.849, {\rxte}
observed {\igr} for $\sim$ 24.5 d, achieving a total exposure of 611
ks ($\simeq$30\% of the whole outburst; ObsId P94041). The 2--60 keV
light curve extracted from the PCU2 of the PCA \citep{Brd93,Jah06}
data is plotted in Fig.\ref{fig:lc}. Background has been subtracted
using the {\it faint} model (pca\_bkgd\_cmfaintl7\_eMv20051128.mdl),
regardless during the outburst the source crosses the threshold of 40
c/s/PCU above which a {\it bright} model should be used. Such a choice
was made to avoid the unphysical flux discontinuity that would take
place when switching from one model to the other. In Fig.\ref{fig:lc},
filled circles represent observations pointed to the {\igr} position
(ObsId P94041), while crosses refer to pointings in the direction of
the nearby source, {\xte} (ObsId P94042). These two sources are 19.67
arcmin away, and there is an overlap of nearly 80\% between the
respective PCA pointings (the FWHM of the PCA collimator is $\sim$
1$^{\circ}$). {\it RXTE} observations are subject to the contribution
of both sources, no matter which the instrument is actually pointed
to. The detected spin frequency is therefore used to discriminate
which source emits the observed flux.  As during the first observation
reported in Fig.\ref{fig:lc} the 245 Hz periodicity is detected, it
can be attributed to {\igr} even if the instrument was pointing
towards {\xte}. The overlap between {\rxte} observations of these two
sources becomes even more evident when looking at the late part of the
outburst. Around 2009 Oct 8 (Day 26 according to the scale used in
Fig.\ref{fig:lc}, where time is reported in days since 2009 Sep 12.0),
the X-ray flux shows an increase of more than 70\% with respect to the
previous emission.  Most importantly, 435 Hz pulsations are detected,
clearly indicating a renewed activity from {\xte}, as it has also been
confirmed by a narrow pointed {\it Swift} XRT observation
\citep{Mrk09b}. Observations performed between 2009 Oct 8 and Oct 10
can be therefore safely attributed to a dim outburst of {\xte},
similar to those already displayed in 2005 March \citep{Grb05} and
2007 April \citep{Fal07}.

The fact that {\igr} is in the Galactic Bulge not only forces to deal
with a crowded field, but also with the Galactic Ridge emission. It is
in fact easy to see from Fig.\ref{fig:lc} that the PCU2 count-rate
stays at a level of $\sim$ 6.5 c/s even after the {\xte} activity
episode is over. Emission at the same flux level was observed also at
the end of the 2002 outburst of {\xte} (GP05) and owes to the Galactic
ridge. This emission  introduces an additional background to
all the PCA observations. This is clearly indicated by the prominent
6.6 keV iron line that appears in PCA spectra when this additional
background is not accounted for \citep{Mrk09a}. As we show in
Sec.\ref{sec:rxtexmm}, to subtract this background is of key
importance to ensure agreement between the EPIC-pn and RXTE-PCA
spectral data.  To estimate a reliable spectrum of this additional
background we therefore summed up all {\rxte} data from 2009 Oct
11.400 to 22.745 (for an exposure of 87.8 ks). In the following we
refer to this additional background as the Galactic ridge emission.

 In order to extract a PCA spectrum, we have considered only data
taken by the PCU2, that is the only always switched on during the {\rxte}
observations overlapping with the {\xmm} pointing.
Data processing has also been restricted to its top xenon layer, as
it is the less affected by the instrumental background. Only photons
in the 3--50 keV band have been retained as this band is best
calibrated. Data taken by the High Energy X-ray Timing Experiment
\citep[HEXTE,][]{Rth98} in the 35--200 keV band has been extracted
considering the cluster B, which was the only one to perform rocking
to estimate the background, at the time the observation took place.

\begin{table}
\caption{Best fitting orbital and spin parameters of {\igr} as
  observed by {\xmm}. Errors in parentheses are evaluated at 1$\sigma$
  confidence level, while upper limits are given at a 3$\sigma$
  level.}
\label{table:spin}
\centering
\renewcommand{\footnoterule}{}  
\begin{tabular}{lr}
\hline 
a $\sin{i}$/c (lt-s) & 0.275196(4)   \\
P$_{orb}$ (s) & 12487.51(2)\\
T$^*$ (MJD) & 55094.9695351(7)\\
e & $<8\times10^{-5}$\\
f(M$_1$,M$_2$,i) (M$_{\odot}$) & $1.07086(5)\times10^{-3}$\\
\hline
$\nu$ (Hz) & 244.8339512(1) \\
$\dot{\nu}$ (Hz/s) & $<3\times10^{-11}$\\
\hline
\end{tabular}
\end{table}

\section{The ''persistent'' emission}

\subsection{The pulse profile}
\label{sec:pulse}

In this paper the pulsations shown by {\igr} are analysed using only
the {\xmm} data. A temporal analysis based on {\rxte} data is instead
included in a companion paper (Riggio et al., 2010, submitted).  In
  order to perform a timing analysis we retain all the 0.3--12 keV
  energy interval covered by the EPIC-pn, and report the photons
  arrival time to the Solar System barycentre using the SAS task {\it
    barycen},  considering the {\it Chandra} position given by
  \citet{Now09}.  The 244.8 Hz periodicity is clearly detected
throughout the {\xmm} observation. To estimate the spin and orbital
parameters we focus on the {\it persistent} emission. We first fold
100 s long intervals around the best estimate of the spin frequency
given by \citet{Rgg09}. The pulse profiles thus obtained are modelled
with a sum of harmonic functions:
\begin{equation}
\label{eq:harmonics}
F(\phi)=A_0\left\{1+\sum_k A_k cos[ k(\phi-\phi_k)]\right\}
\end{equation}
where $A_0$ is the average countrate, $\phi=2\pi\nu (t-T_0)$ is the
rotational phase, and $A_k$ and $\phi_k$ are the fractional amplitude
and phase of the k-th harmonic, respectively. Two harmonics are enough
to model pulse profiles obtained over 100s intervals. The relevant
orbital parameters (the semi-major axis of the NS orbit,
$x=a\sin{i}/c$, the orbital period, $P_{orb}$, the epoch of passage at
the ascending node of the orbit, $T^*$, and the eccentricity $e$) are
estimated fitting the time delays affecting the first harmonic phases
with respect to a constant frequency model (see Eq.(3) of
\citealt{Pap07}, and references therein).  The observed time delays,
together with the residuals with respect to the best fitting model,
are plotted in Fig.\ref{fig:orbit}, while the orbital parameters we
obtain are listed in Table \ref{table:spin}. They are perfectly
compatible with those estimated by Riggio et al. (2010, submitted)
considering the RXTE coverage of the outburst, even if the
uncertainties that affect them are naturally larger because of the
shorter exposure they are calculated over.

\begin{figure}
 \includegraphics[angle=270.0,width=\columnwidth]{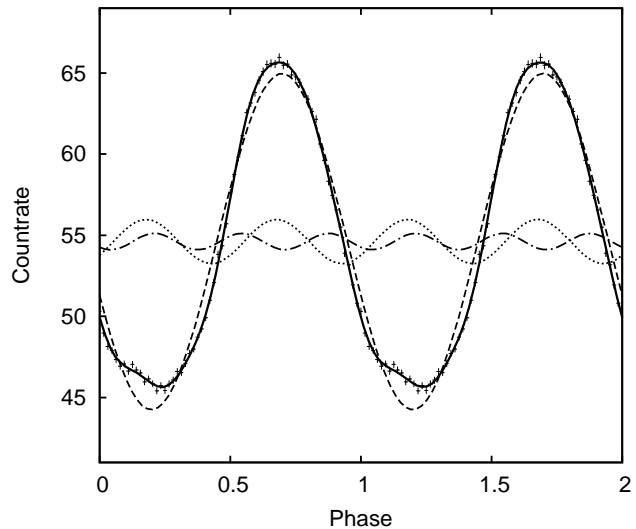}
 \caption{Pulse profile obtained by folding the 0.3--12keV EPIC-pn
   time series in 64 bins around the best spin frequency estimate,
   $\nu=244.8339512$, after having corrected data for the orbital
   motion. The solid line is the best fitting five harmonics function
   described in text. The first (dashed line), second (dotted line)
   and third (dash-dotted) harmonic are also over-plotted. An offset
   equal to the average count rate have been added to the harmonics
   count-rate for graphical purposes. Two cycles are shown for clarity.
 }
\label{fig:profile}
\end{figure}

After the photon arrival times have been corrected for the orbital
motion of the pulsar, we folded data in 16 phase bins over longer time
intervals (1000s), and derived more stringent measures of the phases
and of their temporal evolution.  The statistical errors on the phases
obtained fitting the profiles so obtained are summed in quadrature
with the uncertainty introduced by the errors affecting the orbital
parameters (see Eq.(4) in \citealt{Pap07}).  As there are no
significant differences between the estimates we obtain with the two
harmonic components, we present here only the results based on the
fundamental one.  The phase evolution can be successfully modelled by
a constant frequency ($\chi^2_{red}=1.02$ for 68 d.o.f.), whose
estimate is given in Table \ref{table:spin}. The quoted error is
evaluated considering also the uncertainty introduced by the
positional error-box (see \citealt{Bur07}).  The introduction of a
quadratic term, possibly reflecting a spin evolution of the source, is
not significant, and it is possible to derive a 3$\sigma$ upper limit
on the spin frequency derivative of $1\times10^{-11}$ Hz/s.

\begin{figure}
 \includegraphics[angle=270.0,width=\columnwidth]{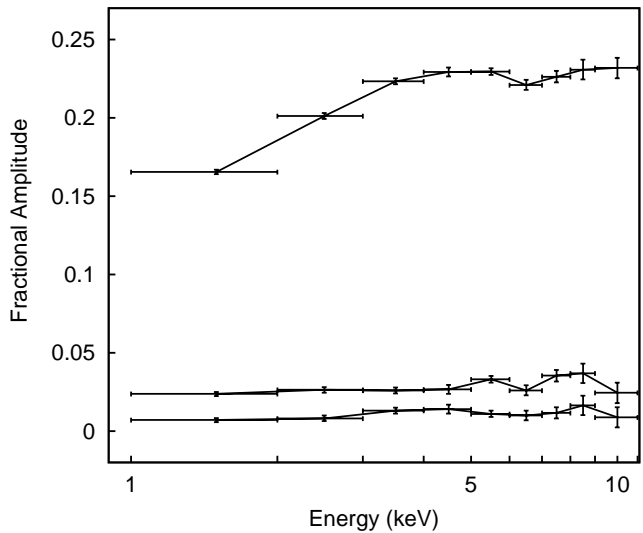}
 \caption{Variation with energy of the fractional amplitude of the first (top), second (middle) and third (bottom) harmonic.
 }
\label{fig:ampiezze}
\end{figure}

Folding over the entire observation length the 0.3--12 keV EPIC-pn
time series we obtain the average pulse profile plotted in
Fig.\ref{fig:profile}.  Five harmonics are needed to successfully fit
(i.e. $\chi^2/$dof=53.1/54) the pulse profile using
Eq.(\ref{eq:harmonics}), while AMSPs pulse profiles are generally
fitted by just two components\footnote{We note that more than two
    harmonics has been detected during a subset of observations of
    {\saxj} \citep{Hrt08} and XTE J1807--294 \citep{Pat09b}}. Such a
complexity is unveiled probably thanks to the high statistics and low
background granted by the EPIC-pn, and, most importantly, to the
larger pulse fraction {\igr} shows with respect to the other objects
of this class. Accounting for the average background count-rate
(2.225(6) in the 0.3--12 keV band), the fractional amplitudes of these
harmonic components are in fact, $20.21(8)\%$, $2.66(8)\%$,
$0.98(8)\%$, $0.39(8)\%$ and $0.33(8)\%$, respectively, where the
errors in parentheses are quoted at 1$\sigma$ confidence level. The
total RMS pulsed fraction can be estimated as $(\sum_k
A_k^2/2)^{1/2}=14.4(3) \%$ RMS.

\begin{figure}
 \includegraphics[angle=270.0,width=\columnwidth]{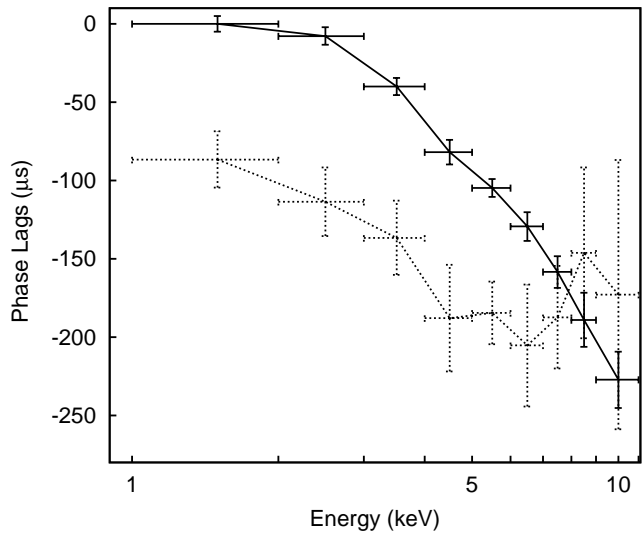}
 \caption{Phases in $\mu s$ of the first (solid line) and second
   harmonic (dotted line) with respect to energy. An common offset has
   been set on phases so that the phase of the fundamental between 1
   and 2 keV is equal to 0.  }
\label{fig:phaselag}
\end{figure}

The pulse of {\igr} also shows strong spectral variability. The
amplitude of the first harmonic increases with energy until it reaches
an approximately constant value of $\sim0.22$ at $\sim 3 keV$.  The
second and third harmonic are instead more regular (see
Fig. \ref{fig:ampiezze}). We show in Sec.\ref{disc:spectrum} how the
decrease of the pulsed fraction can be understood in terms of the
shape of the various components used to model the X-ray spectrum.
Phase lags are also observed (see Fig.\ref{fig:phaselag}). In
particular the phase of the fundamental shows an excursion of
$\sim230\mu s$ (0.06 in phase units) between 1 and 10 keV, with the
pulses at low energy lagging those at higher energies.  Lags are also
shown by the second harmonic, even if their significance is lowered by
the larger error bars affecting these estimates, with respect to those
calculated on the first harmonic. The behaviour with energy of the
second harmonic phases is anyway more regular. While at low energies
the second harmonic phase anticipates the first harmonic, at higher
energies they become comparable. We thus conclude that the overall
pulse shape is energy dependent.

\begin{table*}
\begin{minipage}[t]{170mm}
\caption{ Fitting parameters of the 0.5-11.0 keV combined RGS +
  EPIC--pn spectrum of {\igr}. Models are defined as follows: Model A,
  \texttt{phabs}$\times$\texttt{edge}(\texttt{diskbb}+\texttt{bbodyrad}+\texttt{nthcomp}) with
  kT$_{soft}=$kT$_{BB}$; Model B,
  \texttt{phabs}$\times$\texttt{edge}(\texttt{diskbb}+\texttt{bbodyrad}+\texttt{nthcomp}) with
  kT$_{soft}\neq$kT$_{BB}$; Model C, same as model Model B with the
  addition of a \texttt{diskline} to model the K$\alpha$ iron emission
  feature. The disc outer radius is fixed to $1\times10^5$ R$_g$. All
  these models are evaluated for an electron temperature of the
  Comptonized component fixed to 100 keV. Errors on each parameter are
  quoted at the 90\% confidence level, as for all the spectral
  parameters in the rest of the paper.  }
\label{tab1}
\centering
\renewcommand{\footnoterule}{}  
\begin{tabular}{lrrr}
\hline \hline
Model &  A & B & C\\
\hline
nH ($10^{22}$ cm$^{-2}$)  &$1.10^{+0.07}_{-0.04}$ &$0.98\pm0.04$ & $0.96^{+0.01}_{-0.02}$ \\
$\tau$ O VIII  & $0.22\pm0.08$&$0.21\pm0.07$& $0.20\pm0.07$ \\
kT$_{in}$ (keV) &$0.21^{+0.02}_{-0.01}$ &$0.33^{+0.04}_{-0.02}$&$0.34^{+0.04}_{-0.03}$ \\
R$_{in}(\cos{i})^{1/2}$  ($d_{8}$ km)  & $63^{+28}_{-19}$&$20^{+5}_{-4}$& $17^{+4}_{-5}$ \\
kT$_{BB}$ (keV) & $0.39^{+0.02}_{-0.03}$&$0.62^{+0.03}_{-0.02}$&$0.63^{+0.02}_{-0.03}$ \\
R$_{BB}$ ($d_{8}$ km) & $7\pm1$&$7.1^{+0.5}_{-0.6}$&$6.7^{+0.3}_{-0.4}$ \\
$\Gamma$  &$1.710\pm0.008$ &$2.4\pm0.2$ &$2.1^{+0.2}_{-0.1}$ \\
kT$_{soft}$ (keV)  &=kT$_{BB}$  &$1.5\pm0.01$ & $1.36^{+0.05}_{-0.17}$ \\
\hline
E$_{Fe}$ (keV) & --- & --- & $6.72^{+0.24}_{-0.09}$\\
$\beta$ &--- &--- &$-4.1^{+0.8}_{-2.9}$\\
R$_{in}$ ($GM/c^2$)  &--- &---& $37^{+31}_{-9}$\\
i $(^{\circ})$ &--- &--- &$>37$\\
EW (eV) &---&---&$43.9\pm0.06$\\
\hline
$\chi^2_{red}$  &$ 2008.1/1598$ &$1775.8/1597$& $1740.7/1592$\\
\hline
\end{tabular}
\end{minipage}
\end{table*}

\subsection{The {\xmm} spectrum}
\label{sec:xmmspectrum}

We first analyse the {\xmm} spectrum, considering data taken by the
two RGS (0.5--2.0 keV) and by the EPIC-pn (1.4--11.0 keV).  EPIC-pn
  data below 1.4 keV are discarded as they show a clear soft excess
  with respect to RGS data, regardless of the model used to fit the
  spectrum. Such an excess was already noticed by, e.g.,
  \citet{Brn05,Iar09,Dai09,Pap09,Dai10}, analysing observations
  performed by the EPIC-pn in timing mode. The absence of such a
  feature in RGS data indicates how it is probably of calibration
  origin, hence the choice to retain only data taken by the EPIC-on in
  the 1.4--11.0 keV energy interval, for the purposes of a spectral
  analysis. The relative normalisation of the RGS1 and RGS2 with
respect to the EPIC-pn camera are left free. We find RGS1/PN=$1.02(1)$
and RGS2/PN=$0.98(1)$, regardless of the particular spectral model
considered. The numbers quoted in parentheses are the errors at 90\%
confidence level, level at which the uncertainty on any spectral
parameter is quoted in this paper.

  The 0.5--11 keV spectrum of {\igr} is energetically dominated by a
  power law, which is easy to interpret as the realization of a
  broader Comptonized emission on a limited bandwidth.  The first
  Comptonization model we consider is \texttt{nthcomp}
  \citep{Zdz96,Zyc99}. It describes a thermal Comptonization spectrum
  in terms of the temperature of the input soft photons (kT$_{soft}$),
  the temperature of the hot comptonizing electrons (kT$_{e}$) and by
  an asymptotic power law index ($\Gamma$) related to kT$_{e}$ and to
  the medium optical depth $\tau$ through the relation:
\begin{equation}
\label{eq:gamma}
\Gamma=\left[ \frac{9}{4}+\frac{1}{\left(\frac{kT_e}{m_ec^2}\right)\tau\left(1+\frac{\tau}{3}\right)}\right]^{1/2}-\frac{1}{2}
\end{equation} (see e.g. \citealt{LigZdz87}).
 As no high energy cut-off appears in the {\xmm} spectrum, we fix
 kT$_{e}$ to an arbitrary value of 100 keV. Such a choice does not
 affect significantly the results obtained on the {\xmm} dataset
 alone. Interstellar absorption is treated with the \texttt{phabs}
 model, the abundances are fixed at the values of \citet{AndGre89},
 and cross sections are taken from \citet{BalMcC92}, and modifications
 after \citet{YanSad98}. Data show no significant deviation from solar
 abundances. The \texttt{nthcomp} component alone, poorly reproduces
 the observed spectrum ($\chi^2_{red}=1.68$ for 1604 d.o.f.).
 Residuals around 1.8 and 2.2 keV are evident.  As there is no sign of
 similar features in the RGS spectra, we interpret them in terms of an
 incorrect calibration of the instrumental Si and Au edges that
 frequently affect EPIC-pn spectra. An improvement of
 $\Delta\chi^2=151$ is obtained modelling these residuals with narrow
 Gaussian absorption features.

As a soft excess is present, we add two thermal components to the
model, accordingly to the results previously obtained modelling {\xmm}
spectra of AMSP (GP05, \citealt{Pap09}). We model the softer component
as disc emission (\texttt{diskbb}), and the hotter as a single
temperature blackbody (\texttt{bbodyrad}). As the normalisation of the
latter component is smaller than $\approx$ 10 km, we interpret it as
thermal emission arising from the neutron star surface.  Assuming that
it is this component that provides the seed photons for
Comptonization, we tie kT$_{soft}$ to the temperature of the
blackbody.  The addition of these two thermal components definitely
improves the fit, as the $\chi^2$ decreases of 39 and 486, for the
addition of two and one degree of freedom, respectively.  An
absorption edge at $0.88\pm0.01$ keV is also clearly detected in the
RGS data, with an absorption depth of $0.22\pm0.06$. We identify it as
an absorption edge of O VIII (E$_{\:OVIII}=0.871$ keV). The edge is
quite sharp, as an upper limit of $34$ eV is found on its width, if a
smeared edge (\texttt{smedge}) is used to model it. The best fitting
parameters of this model (named as model A) are listed in the left
column of Table \ref{tab1}.

The final reduced chi squared is anyway still large,
$\chi^2_{red}$=1.26(1598 d.o.f.).  Disentangling the temperature of
the seed photons of the Comptonized component from the observed
blackbody (model B, see Table \ref{tab1}), significantly improves the
fit ($\Delta\chi^2=232$ for the addition of just one d.o.f.). Data
thus favour a hotter and smaller region to produce the seed photons of
the Comptonized component, than that of the observed
\texttt{blackbody}.

The presence of residuals around 6.6 keV (see Fig.\ref{fig:residui}
where EPIC-pn residuals with respect to a simple power law are plotted
in the bottom panel) suggests the presence of an iron K$\alpha$
emission line. Adding a Gaussian centred at $E_{Fe}=6.65 \pm 0.25$ keV
improves the $\chi^2$ by 23 (for three additional d.o.f.). The width
of the feature ($\sigma>0.56$ keV) indicates it could be produced from
reflection of the NS hard emission on the geometrically thin,
optically thick, accretion disc. In this context, the feature is
broadened by the relativistic motion of the reflecting plasma in the
inner parts of the accretion disc, where the space-time is bent by the
gravitational influence of the compact object.  We use a
\texttt{diskline} model to account for relativistic effects
\citep{Fab89}, obtaining an improvement of $\Delta\chi^2=13$ for 3
d.o.f. added. The \texttt{diskline} model describes the line shape in
terms of the size of the illuminated disc (that is, of its inner
radius, $R_{in}$, and of its outer radius, $R_{out}$), of the index of
the radial dependence of the line emissivity, $\beta$, and of the
inclination of the system, $i$. As $R_{out}$ is poorly constrained by
the available statistics, we fix it to an arbitrary value of $10^5$
R$_g$ (where $R_g=GM_1/c^2$ is the NS gravitational radius, and $M_1$
the mass of the compact object), of the order of the circularization
radius for a system like {\igr} \citep[see e.g.][]{FrnKngRai02}. The
parameters we obtain are listed in the rightmost column of Table
\ref{tab1} (model C), and are compatible with the assumption that the
line is emitted by plasma illuminated by the NS emission, and rotating
in the accretion disc. The energy of the transition indicates mildly
to highly ionised iron. As we show in the next section, these
indications are supported also by the modelling of disc reflection on
a broader energy range. The feature stands at 3.5$\sigma$ above the
continuum, and the F probability that the $\chi^2$ improvement when
switching from model B to model C is due to chance is
$<10^{-5}$. However, the use of an F-Test to test the presence of
additional components has been strongly discouraged by \citet[][but
  see also \citealt{Stw09}]{Prt02}. We have then simulated 100 spectra
starting from the best fit parameters of model B, and fitted them
using both model B and C. In none of the cases we have noticed a
$\chi^2$ improvement, or an F statistics value, larger than the one we
obtain from modelling real data. We therefore exclude at more than 99
per cent confidence level that the $\chi^2$ improvement obtained with
model C with respect to model B is due to chance.

 Model C is the best model we have found to fit the {\xmm}
 dataset. However the $\chi^2$ corresponding to this model is
 relatively large and the probability that we observe by chance a
 $\chi^2$ equal or larger than the value we obtain, if the model is
 correct, is 0.5 per cent. This makes the model only barely
 acceptable. The average scatter of data points with respect to this
 model is anyway compatible with the accuracy of the effective area
 calibration of the EPIC-pn, while operated in fast modes (quoted to
 be better than 5\% from an analysis of the 1.5--3 keV band, see
 http://xmm2.esac.esa.int/docs/documents/CAL-TN-0083.pdf).  Since the
 distribution of residuals does not show any systematic trend, but are
 instead randomly distributed around what is predicted by our best fit
 model, we conclude that the $\chi^2$ we obtain is affected by
 uncertainties in the instrument calibration, and/or by the possible
 presence of unresolved and unfitted features. We therefore retain
 model C as a reliable description of IGR J17511-3057 spectrum in the
 considered bandwidth.

\subsection{A simultaneous {\rxte}--{\xmm} spectrum}
\label{sec:rxtexmm}

The spectrum of {\igr} is dominated by a power-law like component
which we have interpreted in terms of Comptonization, with a cut-off
temperature beyond the energy band covered by the EPIC-pn. Moreover,
the detection of a broadened K$\alpha$ iron emission feature indicates
how disc reflection may be important. To better assess these issues,
we take advantage of the observations performed by {\rxte} (3--200
keV) overlapping with the {\xmm} pointing. We thus added to the {\xmm}
data presented in the previous section, the PCA (3--50 keV) and the
HEXTE (35--200 keV) spectra. A systematic error of 0.5\% has been
added to the PCA points according to the guidelines stated in the
description of the latest PCA response matrix
generator \footnote{http://www.universe.nasa.gov/xrays/programs/rxte/pca/doc/\\/rmf/pcarmf-11.7/}.

\begin{figure}
\begin{centering}
 \includegraphics[angle=270,width=\columnwidth]{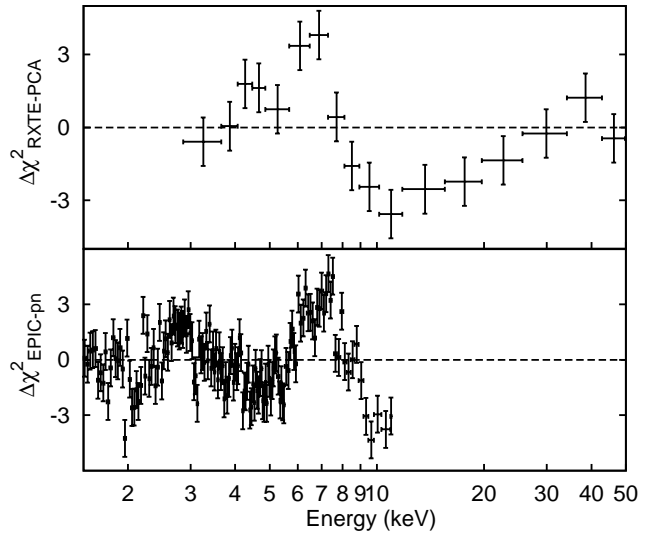}
 \caption{Residuals of the RXTE-PCA (top panel) and of the EPIC-pn
   (bottom panel ) with respect to an absorbed power law model. The
   RXTE-PCA points are subtracted from an additional background
   representing the emission of the Galactic ridge. }
\label{fig:residui}
\end{centering}
\end{figure}

In order to check the inter calibration between the EPIC-pn and
RXTE-PCA spectra, we simultaneously fit these two spectra with an
absorbed power law. The RXTE-PCA spectrum shows large swings up to
$\sim$ 10$\sigma$ with respect to EPIC-pn points, with a clear soft
excess below 5 keV and a much more prominent emission feature at
$\sim$ 6.6 keV. In order to investigate if such a discrepancy could be
due to the contamination of the Galactic ridge emission, we subtracted
from PCA data the Galactic ridge spectrum (see
Sec.\ref{sec:rxte})\footnote{No additional background was considered
  for HEXTE data, as the Galactic ridge emission above 15 keV is
  minimal.}. A great improvement ($\Delta\chi^2=376.33$ for 710
d.o.f.)  is achieved with respect to the unsubtracted data. Most
importantly, the residuals of the EPIC-pn and the PCA points are
distributed in the same way after the subtraction of the Galactic
ridge contribution to the PCA spectrum (see
Fig. \ref{fig:residui}). Such a result gives us confidence about the
reliability of the combined {\xmm}--{\rxte} spectral modelling we
describe in the following.  The relative normalisations between the
various instruments and the EPIC-pn are left free.  The same
normalisations quoted in the previous section have been found for the
RGS, while PCA/PN and HEXTE/PN vary in the ranges $1.24$--$1.25$ and
$0.6$--$0.8$, respectively, depending on the particular model used. We
note that the values taken by the normalisation factors for different
spectral models are all compatible with each other within the
respective errors (that are typically $0.006$ and $0.1$ for PCA/PN and
HEXTE/PN, respectively).

\begin{table*}
\begin{minipage}[t]{170mm}
\caption{
Best-fitting parameters of the 0.5--200 keV spectrum composed
  of the RGS (0.5--2.0 keV), EPIC-pn (1.4--11 keV), PCA (3.0--50.0
  keV) and HEXTE (35--200 keV) spectra. The continuum models are
  defined as follows:  (i) Model B, \texttt{phabs$\times$edge(diskbb+bbodyrad +nthcomp)}; (ii) Model B-Refl, \texttt{phabs$\times$edge(diskbb+bbodyrad+nthcomp+diskline+rdblur$\times$pexriv)}; (iii) Model PS, \texttt{phabs$\times$edge(diskbb+bbodyrad+compps)}; (iv) Model PS-Refl, \texttt{phabs$\times$edge(diskbb+bbodyrad+compps+diskline+rdblur$\times$pexriv)}.   All the fluxes calculated are unabsorbed.}
\label{tab2}
\centering \renewcommand{\footnoterule}{} 
\begin{tabular}{lrrrr}
\hline \hline
Model &B & B-Refl & PS & PS-Refl \\
\hline
nH ($10^{22}$ cm$^{-2}$)&$1.04^{+0.04}_{-0.05}$&$0.98_{-0.05}^{+0.04}$ &$1.00^{+0.03}_{-0.04}$&$0.95^{+0.02}_{-0.03}$  \\
$\tau$ O VIII   &$0.22^{+0.07}_{-0.08}$ &$0.14\pm0.08$ &$0.22\pm0.07$&$0.19\pm0.07$ \\
\hline
kT$_{in}$ (keV)  &$0.27^{+0.03}_{-0.02}$&$0.32^{+0.05}_{-0.02}$&$0.31^{+0.03}_{-0.02} $&$0.36\pm0.02$   \\
R$_{in}\sqrt{\cos{i}}$  ($d_{8}$ km) &$33\pm9$&$19\pm8$&$24^{+7}_{-4}$&$15^{+1}_{-2}$  \\
F$_{Disc}^{bol}$($\times10^{-9}$erg/cm$^2$/s)\footnote{No error estimate is given as the unabsorbed disk flux is evaluated down to 0.01 keV, that is below the lower energy bound covered by data. } &$0.20$  &$0.13$ &$0.17$ &$0.14$  \\
\hline
kT$_{BB}$ (keV)         &$0.53^{+0.02}_{-0.01}$& $0.60\pm0.04$ &$0.60\pm0.02$ &$0.64^{+0.01}_{-0.02}$ \\
R$_{BB}$ ($d_{8}$ km)  &$8.7^{+0.6}_{-0.7}$&$6.8\pm0.6$&$7.6\pm0.5$&$6.3^{+0.3}_{-0.1}$ \\
F$_{BB}^{0.5-150}$($\times10^{-9}$erg/cm$^2$/s) &$0.10\pm0.02$ &$0.10\pm0.02$ &$0.10\pm0.01$ & $0.11\pm0.01$\\
\hline
$\Gamma$              &$1.86^{+0.03}_{-0.02}$& $1.78\pm0.02$  & ---   & ----  \\
kT$_{el}$(keV)            &$100$(fixed)&$100$(fixed)   &$75^{+11}_{-10}$      &  $51^{+6}_{-4}$ \\
$\tau$  \footnote{for models B and B-Refl the range of optical depths has been evaluated from the best-fitting values of $\Gamma$ obtained fixing kT$_{e}$ at values in the range 50--200 keV, and using Eq.(\ref{eq:gamma}).} 
      &$0.6$--$1.9$  &$1.1$--$2.0$ &$1.18^{+0.08}_{-0.07}$ &$1.34^{+0.03}_{-0.06}$\\
i$_{SLAB}(^{\circ})$  &---&--- &$65.0(fixed)$&$57.0^{+11}_{-7}$ \\
kT$_{soft}$          &$1.01^{+0.04}_{-0.05}$  &$1.1\pm0.1$   &$1.47\pm0.06$   &$1.37^{+0.01}_{0.02}$  \\
R$_{soft}$($d_{8}$ km)\footnote{for models B and B-refl it has been evaluated assuming $F_C=(e^y-1)F_{soft}$ (see Eq.\ref{eq:rsoft}).} &$4.2$--$5.3$ &$3.2$--$4.7$  &$3.3^{+0.3}_{-0.2}$&$3.4^{+0.1}_{-0.4}$ \\
F$_{C}^{(0.5-150)}$($\times10^{-9}$erg/cm$^2$/s) &$1.00\pm0.04$  &$1.04\pm0.07$ &$1.30\pm0.08$&$1.00\pm0.05$ \\
\hline
$\log\xi$ &--- &$3.4\pm0.5$ &---&$3.0^{+0.4}_{-0.2}$ \\
$\Omega/2\pi$ &--- &$0.18\pm0.05$ &--- &$0.14^{+0.10}_{-0.01}$ \\
F$_{Refl}^{(0.5-150)}$($\times10^{-9}$erg/cm$^2$/s)&---&$0.07\pm0.02$&---&$0.07\pm0.03$\\
\hline
E$_{Fe}$ (keV) & --- & $6.8^{+0.2}_{-0.1}$ & --- & $6.82^{+0.09}_{-0.11}$ \\
$\beta$ &--- &$-6_{-6}^{+3}$ &---& $-5.5^{+0.8}_{-4.9}$\\
R$_{in}$ ($GM/c^2$) &--- &$22^{+9}_{-5}$&--- & $27^{+6}_{-9}$\\
i $(^{\circ})$ &--- &$41_{-5}^{+8}$ &---& $48^{+6}_{-10}$\\
EW (eV) & --- &$35\pm1$ & --- &$40.9\pm0.7$ \\
\hline
F$_{unabs}^{0.5-150}$($\times10^{-9}$erg/cm$^2$/s)&$1.19\pm0.04$ &$1.28\pm0.08$ &$1.52\pm0.08$& $1.28\pm0.06$\\
$\chi^2_{red}$ &$1993.1/1721$  &$1838.1/1713$&$1912.7/1720$&$1847.7/1712$ \\
\hline
\hline
\end{tabular}
\end{minipage}
\end{table*}

\begin{figure*}
\begin{minipage}[t]{170mm}
\centering
 \includegraphics[angle=270,width=\columnwidth]{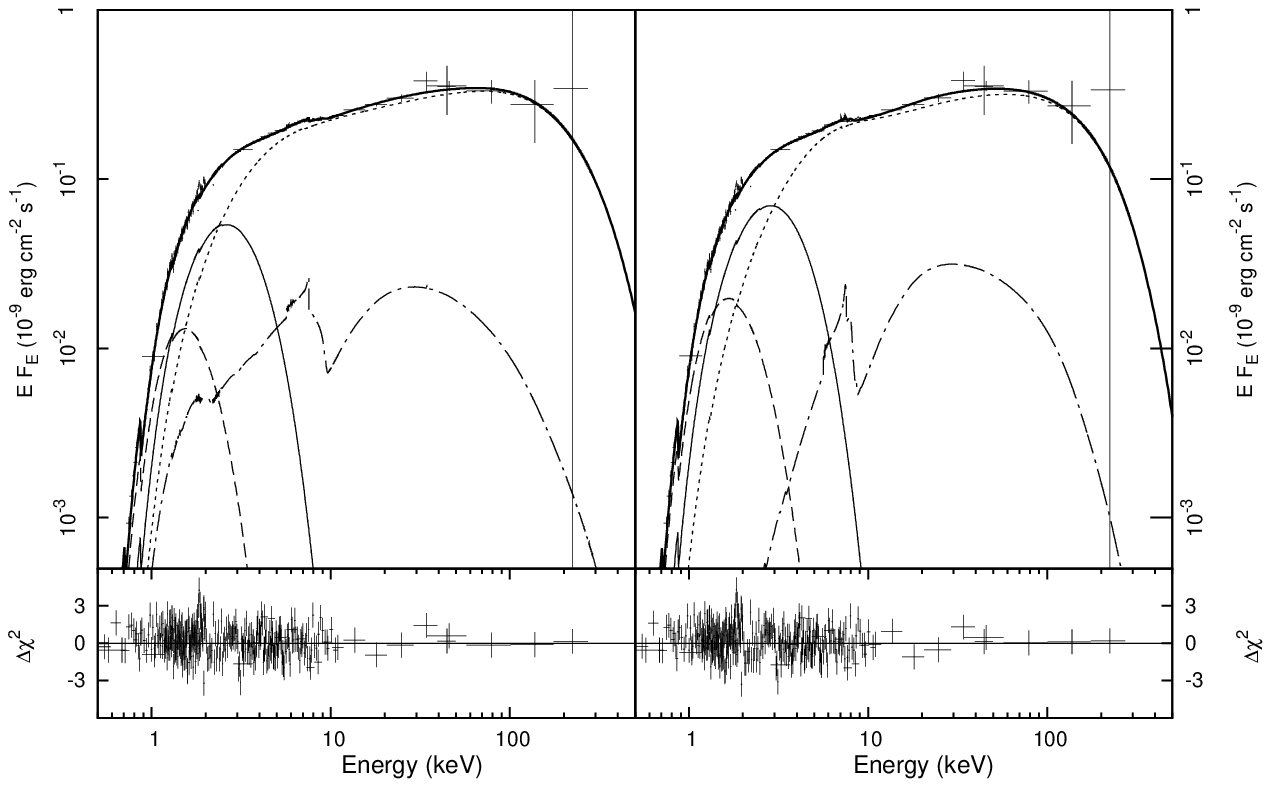}
 \caption{Unfolded 0.5--200 keV spectrum of {\igr} convolved with the
   model B+refl (left) and PS+Refl (right; see caption of Table
   \ref{tab2} for a description), together with residuals in units of
   sigmas in the respective bottom panels. The overall model (thick
   solid line), the \texttt{diskbb} (dashed line), the
   \texttt{blackbody} (solid thin line), the Comptonized spectrum
   (dotted line), and the reflected spectrum (continuum and
   \texttt{diskline}, dash-dotted line) are also overplotted.  Data
   have been re-binned for graphical purposes. }
\label{fig:compps}
\end{minipage}
\end{figure*}

As expected, the analysis of the combined datasets gives similar
results than those presented in the previous section for what concerns
the low energy part of the spectrum. Two soft components are requested
to model the observed data, in addition to an $\alpha=1.56\pm0.04$
power law, cut off at an energy $E=51^{+16}_{-9}$ keV. Using
\texttt{nthcomp} to model the hard emission we find that, if the
observed \texttt{blackbody} provides the seed photons upscattered in
the Comptonizing medium ($kT_{BB}=kT_{soft}=0.34^{+0.03}_{-0.04}$
keV), chi squared is large ($\chi^2_{red}=1.26$ for 1722 d.o.f.).
Moreover, a region of size larger than the one measured
($R_{BB}=8.6\pm2.7$ d$_8$ km, where d$_8$ is the distance to the
source in units of 8 kpc) is needed to provide enough photons for the
observed hard component. Such a radius is evaluated considering that
the flux escaping the Comptonizing medium is $F'=A\times F_{soft}$,
where $F_{soft}$ is the seed flux, and the Compton amplification
factor is $A=e^y$, with $y=(4kT_e/m_ec^2)\times \max{(\tau,\tau^2)}$,
the Compton parameter\footnote{The better this approximation holds the
  lowest the average energy of scattered photon is with respect to the
  electron temperature, see e.g. \citet{RybLig79}.}. As the
\texttt{nthcomp} model only evaluates the flux added by the Compton
process, $F_C$, the energy conservation between the two phases can be
rephrased as $F_C/F_{soft}\simeq e^y-1$. Expressing the seed flux as
$F_{soft}=\sigma T_{soft}^4(R_{soft}/d)^2$, the radius of the area
providing the seed photons has to be:
\begin{equation}
\label{eq:rsoft}
R_{soft}= 2.4 \times 10^5 d_{8} \frac{\sqrt{F_{C}/(e^y-1)}}{(kT_{soft})^2} km,
\end{equation} 
where $kT_{soft}$ is expressed in keV.  As using \texttt{nthcomp} no
significant cut-off is found, we are forced to measure the
best-fitting values of $\Gamma$ fixing the electron temperature at
values in the range 50--200 keV, and then evaluate $\tau$ and $y$
using Eq.(\ref{eq:gamma}). Estimating the flux in the \texttt{nthcomp}
component in the 0.5--150 keV energy range, the radius we obtain from
Eq.(\ref{eq:rsoft}) is $R_{soft}\simgt30$ d$_8$ km, which is at least
a factor $\sim 3$ larger than the normalisation of the observed
\texttt{blackbody} component. We thus conclude that the observed
\texttt{blackbody} is too cold to be the source of the seed photons.

Letting kT$_{soft} \neq$ kT$_{BB}$ improves the modelling by
$\Delta\chi^2=186$ for the addition of one d.o.f. (see the leftmost
column of Table \ref{tab2} for the best fitting parameters of this
model, named as B). As the electron temperature is not constrained by
the model, we present the best fitting parameters for kT$_{el}$= 100
keV. Using Eq.(\ref{eq:rsoft}) to evaluate the radius of the unseen
thermal component providing the seed photons up-scattered in the
Comptonizing medium, we find $R_{soft}\simeq$4--5 d$_8$ km.
This radius is compatible with the expected size of the hotspots on
the NS surface.

The presence of an iron K$\alpha$ emission line in the {\xmm} spectrum
indicates a significant presence of reflection. To model Compton
reflection from an ionised disc we use the \texttt{pexriv} model
\citep{MgdZdz95}. The shape of the power law that describes the
illuminating flux is fixed using the index $\Gamma$ of the
\texttt{nthcomp} component.  The disc temperature is kept fixed at a
value of $1\times 10^6$ K, the disc inclination at the value indicated
by the iron line modelling, and abundances to solar values. The
relativity effects expected for a Keplerian accretion disc rotating
around a compact object are taken into account with the smearing
kernel \texttt{rdblur} \citep{Fab89}.  To include also the most
important bound-bound transition expected (Fe K$\alpha$) in the
EPIC-pn range, we add a \texttt{diskline} with a line energy free to
vary in the 6.4--6.97 keV interval. The outer radius of the
illuminated disc is fixed at $10^5$ R$_g$, as the fit is rather
insensitive for variations of this parameter in a wide range of
values.  The addition of a reflection component is highly significant,
as the $\chi^2$ improves by 155 for the addition of 9 parameters (see
the model BRefl in Table \ref{tab2} for the best fitting parameters).

As the Comptonization model \texttt{nthcomp} is not able to constrain
the electron temperature, the assumption we have made (kT$_{el}$=100
keV) has potentially an impact on the amount of reflection needed to
model the spectrum. The value of kT$_{el}$ also influences the
determination of the seed-photon region (see Eq. \ref{eq:rsoft}). The
results obtained considering both the 0.5--11 keV and the 0.5--200 keV
energy bands indicate that the seed photons come from a region which
is hotter and smaller than the \texttt{blackbody}, and not directly
observed. It is then desirable to include in the model those seed
photons that are not scattered in the $\tau\approx1$ Comptonizing
medium, as \texttt{nthcomp} only accounts for up-scattered photons.
For all these reason, as well as to check the dependence of the
measured parameters on the Comptonization model chosen, we substitute
\texttt{nthcomp} with \texttt{compps} \citep{PouSve96}. This model has
in fact a number of advantages with respect to the \texttt{nthcomp}
model: (i) it enforces the energy balance between the Comptonizing
plasma and the region that provides the soft photons; (ii) it
evaluates the Comptonized spectrum by solving numerically the
radiative transfer equation, for different scattering orders, and for
different geometries of the comptonizing medium.  It thus includes
those seed photons that go through the hot medium unscattered.  We
consider in the following a slab geometry; (iii) it uses the
\texttt{pexriv} kernel to model disc reflection of the hard photons
only. The reflected spectrum thus implicitly takes into account the
decrease of photons incident on the disc at low ($\simlt 2$ keV)
energies, while using the model B-Refl the incident spectrum is
approximated as a power law extending to low energies.

When reflection is not included in the model we obtain the results
listed in Table \ref{tab2} (Model PS). The continuum parameters do not
vary much when using this Comptonization model. In particular, the
size of the region that provides the seed photons confirms the results
previously obtained.  Using this model, we are also able to constrain
the electron temperature of the hot medium, in agreement with the cut
off energy found using a simple power law to model the hard emission.

The property (ii) quoted above introduces a dependence of the emergent
spectrum on the angle i$_{slab}$ between the slab normal and the line
of sight.  Keeping $i_{slab}$ as a free parameter in spectral fitting
yields values $\simgt80^{\circ}$, that would imply the presence of
eclipses, which are instead not observed. As we see in the following,
smaller values are obtained when reflection is included into the
model. To calculate the spectrum presented in Table \ref{tab2} we thus
fix $i_{slab}=65^{\circ}$. The tendency of the \texttt{compps} model
to find large values of $i_{slab}$ when reflection is not included,
can be tentatively interpreted as an indirect indication of how
reflection is significant to model the broad band spectrum. As a
matter of fact, without reflection, a larger flux has to be accounted
for by the Comptonized component.  As GP05 pointed out, at moderate
optical depths like the ones we detect in {\igr} ($\tau\approx 1$),
the Comptonization spectrum depends on $i_{slab}$ mainly in terms of
the amount of unscattered photons observed, $\propto
exp(-\tau/cos\;{i_{slab}})$.  For these optical depths the angular
dependence of the scattered photons is indeed rather flat (see the
curves relative to angles $\simlt70^{\circ}$ in Fig. 4 of
\citealt{SunTit85}).  As the spectral shape (hence kT$_{soft}$) is
well constrained by data at low energies, it seems that
\texttt{compps} gives account of the large flux at high energies
assuming that we see the slab at very large angles. In this way only a
very small fraction of the unscattered photons would be observed and
the total amount of seed photons would be as large as requested to
ensure energy balance between the hot and the cold phase.  Besides the
significant improvement of the model when reflection is included (see
rightmost column of Table \ref{tab2}, model PS-Refl), the fact that
such an overestimate of $i_{slab}$ is not needed when the addition of
reflection decreases the flux at high energies owing to
Comptonization, definitely supports its presence.

The parameters of the iron line we find with model PS-Refl are all
compatible with those found when Comptonization is described by
\texttt{nthcomp}. Also the ionisation state and the amplitude of
reflection indicated by the reflection model are consistent within the
errors.  The models B-refl and PS-refl are plotted in
Fig.\ref{fig:compps} together with residuals.  The addition of a
  reflection component and of an iron line to models B and PS gives
  values of the F-statistics $>7$. In order to test if such an
  improvement may be due to statistical fluctuations, we have repeated
  the procedure described in Sec.\ref{sec:xmmspectrum}, simulating 100
  fake spectra using the best fit parameters of the models without
  reflection features. As we have never obtained an F-statistics value
  similar to that quoted above, when the reflection features are added
  to the model, we conclude at more than 99\% confidence level that the
  improvement obtained with their addition is not due to counting
  statistics.

The best fit models presented for the combined {\rxte}-{\xmm} dataset
are only barely acceptable (null hypothesis probability of 1.8\% and
1.2\% for the models B-Refl and PS-Refl, respectively). Similarly to
what has been noted in the previous section, the absence of systematic
trends in the residuals leads us to conclude that the reduced chi
squared is increased by uncertainties in the instruments calibration
and/or by the possible presence of unresolved and unfitted features.
The fact that models obtained from a fit of five different instruments
give indeed fairly satisfactory results seems anyway worthwhile to
note, and gives confidence about their reliability.

\label{sec:xmmspectrum}

\section{The burst emission}
\label{sec:burst}

\begin{figure}
 \includegraphics[angle=270.0,width=\columnwidth]{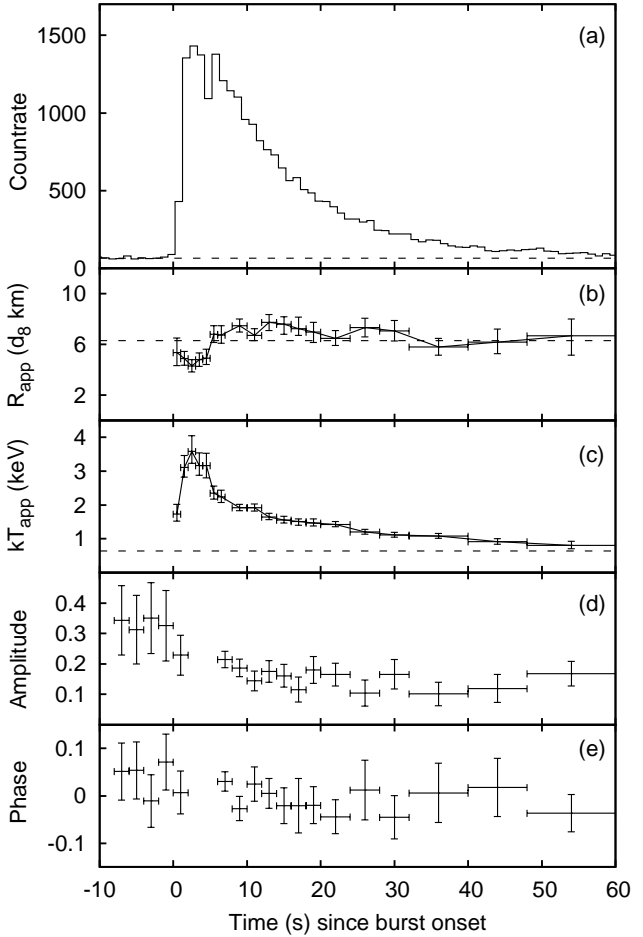}
 \caption{Countrate recorded by the EPIC-pn in the 0.3--12 keV
   interval (a), apparent radius (b) and temperature (c) of the best
   fitting absorbed blackbody, fractional amplitude (d) and phase of
   the pulse first harmonic (e), during the first burst observed by
   the EPIC-pn. Time is given in seconds since the burst onset (MJD
   55095.05025). The values taken by the countrate, the radius and the
   temperature of the \texttt{blackbody} component of model PS-Refl,
   during the {\it persistent} emission are also over-plotted as
   dashed lines for comparison. }
\label{fig:burst1}
\end{figure}

{\igr} exhibited two type I X-ray bursts during the {\xmm} pointing,
sharing similar observational properties (see Table
\ref{table:burst}).  While the rise lasts for $\sim 2$ s, the decay
follows an exponential decay ($\chi^2_{red}=$0.96 and 1.20,
  respectively, over 216 d.o.f.; see the top panel of
Fig.\ref{fig:burst1} for the light curve of the first burst). In order
to study their temporal evolution, we extracted spectra over time
intervals or variable length (from 1s at the beginning to 8s at the
end), subtracting as a background the {\it persistent} spectrum
analysed in the previous section. Pile-up is not of concern, as it
affects energies $\simgt8$ keV for at most few per cent.  We have
anyway checked that the spectrum extracted without the brightest CCD
column gives spectral model parameters compatible with those quoted
hereafter.  We successfully model the burst emission with an absorbed
black body. The absorption column has been varied in the range
indicated by the various models we have used to model the persistent
spectrum.  The evolution of the temperature and of the emission radius
of the first burst, are plotted in panels (b) and (c) of
Fig.\ref{fig:burst1}. Results from the second burst are qualitatively
similar. The burst temperature follows the exponential decay of the
burst flux; the apparent emission radius initially increases but soon
reaches an asymptotic value, suggesting that the critical threshold
for photospheric radius expansion was not reached. In the time
interval 6--30 s, the apparent radius reaches approximately a constant
value, $R_{app}=(7.0\pm0.3) d_8$ km. We restrict to the 6-30 s
interval to estimate this radius as subsequently, the burst flux
becomes comparable with the persistent flux, and systematic errors
could arise due to the subtraction of the persistent emission
\citep{vParLew86}.

In panels (d) and (e) of Fig.\ref{fig:burst1} the amplitude and the
phase of the pulse profile computed on the first harmonic, are also
plotted. Data have been preliminary corrected with the best orbital
solution and folded in 8 phase bins around the spin frequency derived
in Sec.\ref{sec:pulse}. During the first 4s of the burst, the
amplitude decreases to a value compatible with zero. Though, as the
telemetry limit of the EPIC-pn is trespassed at those
countrates, the disappearance of pulsations  may also be due to a dead
time effect. We then discard pulse data referring to the first few
seconds after the burst onset.  However, pulsations are soon recovered
and observed throughout the decaying part of the burst. The amplitude
is roughly comparable with that of {\it persistent} emission, and also
the phase is stable within $\sim$0.1--0.2 cycle. This behaviour suggests
that a  mechanism similar to the one originating the {\it persistent}
pulsations is at work during the type-I X-ray burst (see
Sec.\ref{disc:burst}).

\begin{table}
\begin{minipage}[t]{84mm}
\caption{Parameters of the type I X-ray bursts observed by {\xmm}. Errors in parentheses are evaluated at a 90\% confidence level.}
\label{table:burst}
\centering
\renewcommand{\footnoterule}{}  
\begin{tabular}{lrr}
 & Burst I & Burst II \\
\hline 
T$_{start}$ (MJD) & 55095.05025 & 55095.52363 \\
$\tau$ (s) & 11.4(2) & 11.5(2) \\
F$_{peak}^{\infty}$ $^($\footnote{Bolometric flux in units of $10^{-8}$ erg cm$^{-2}$ s$^{-1}$, evaluated from the best fitting blackbody spectrum at the peak of the outburst.}$^)$ & $4.2\pm1.1$ & $4.3\pm1.1$\\
$\mathcal{F}^{\infty}$ $^($\footnote{Burst fluence in units of  $10^{-7}$ erg cm$^{-2}$}$^)$ & $4.8\pm1.1$ &$4.9\pm1.2$ \\
\hline
\end{tabular}
\end{minipage}
\end{table}

\section{Discussion}

\subsection{The X-ray pulsations}

X-ray coherent pulsations at a frequency $\nu=244.8339512(1)$ Hz are
observed throughout the observation performed by {\xmm}. No spin
frequency evolution is detected at a 3 $\sigma$ upper limit of
$1\times10^{-11}$ Hz s$^{-1}$. Such limit simply reflects the
shortness of the time interval covered by the {\xmm} observation. As a
matter of fact, if a NS spinning at the frequency of {\igr} simply
gains the specific angular momentum of the accreted matter, the
expected spin frequency derivative cannot be larger than
\begin{equation}
\dot{\nu}\simlt 2.4\times10^{-13} I_{45}^{-1} L_{37} R_{10} m_{1,1.4}^{-1/3}\;Hz/s.
\end{equation}
Here $I_{45}$ is the NS moment of inertia in units of $10^{45}$ g
cm$^2$, $m_{1,1.4}$ is the NS mass in units of 1.4 $M_{\odot}$,
$R_{10}$ is the NS radius in units of 10 km, and $L_{37}$ is the
bolometric X-ray luminosity in units of $10^{37}$ erg s$^{-1}$, the
dependence on which have been introduced using the relation $L_X=
GM\dot{M}/R_{NS}$.  Thus, considering the estimate of $L_X$ we give
from spectral modelling ($\simeq 1.6\times 10^{37}$ d$_8^2$ erg
s$^{-1}$) the maximum expected spin frequency derivative is nearly two
order of magnitude lower than the loose upper limit we could set,
given the length of the {\xmm} observation.

The pulsed fraction is among the largest ever observed from an AMSP
(14.4(3) \% RMS). Typical observed values are between 2 and 8 \%, with
sporadic increases up to $\sim$ 10\% especially at the end of the
outburst episodes
\citep{WjnvdK98,Mark02,Gal02,Cam03,Wts05,Gal05,Gal07,Cas08,Alt08,Pat10},
 while \citet{Pat09b} detected a pulse amplitude up to 19\% from XTE
J1807--204.  If the misalignment angle between the magnetic dipole and
the spin axis is small ($\theta\simlt 20^{\circ}$), the observed value
of the pulsed fraction suggests that the inclination is $\simgt
45^{\circ}$ \citep{PouBel06}. Nevertheless, the degree of anisotropy
of the emitted light and the assumed spot shape may significantly
influence the pulse amplitude. A detailed analysis of the pulse
profile is mandatory to derive firm constraints on the spot geometry,
and as it is beyond the scope of this paper, we defer it to another
work. It is worth to note, however, that the presence of a hump in
antiphase with respect to the global maximum possibly suggests that we
see, at least for a fraction of rotational phases, the antipodal spot
\citep[see e.g.][]{Leh09}.

\begin{figure}
\begin{centering}
 \includegraphics[angle=270,width=\columnwidth]{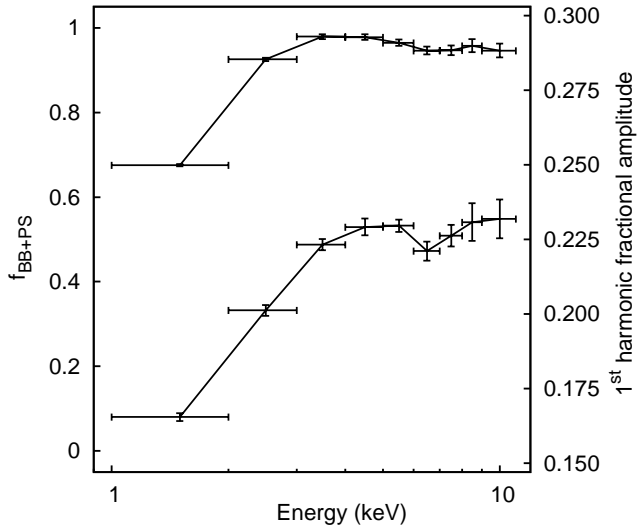}
 \caption{Fractional contribution of the \texttt{bbodyrad} and
   \texttt{compps} components to the EPIC-pn countrate (top curve, left scale), and first harmonic fractional amplitude (bottom curve, right scale). }
\label{fig:ampl_spectrum}
\end{centering}
\end{figure}

The pulse profile of {\igr} shows strong spectral variability,
  similarly to what has been observed from the AMSPs, {\xte} and
  {\saxj}, by GP05 and \citet{Pat09}, respectively. The decrease of
the pulsed fraction at low energies can be completely ascribed to to
the growing influence of the accretion disc emission. To see this, we
plot in Fig.\ref{fig:ampl_spectrum} the fractional contribution of the
\texttt{bbodyrad} and the \texttt{compps} component (Model PS-refl),
to the overall EPIC-pn countrate, $f_{(BB+PS)}$ (top curve, right
scale). The first harmonic fractional amplitude is also plotted
(bottom curve, right scale, same as Fig.\ref{fig:ampiezze}).  The
similarity between the shape of the two curves is striking. The
correlation can also be quantitatively expressed, as the ratio between
the amplitude in the range 1--2 keV and the maximum amplitude
($A_1^{max}=0.232(7)$, E=9--11 keV) is 0.71(2), almost equal to
$f_{(BB+PS)}$ in the same energy band (0.675(3)). The energy
dependence of the pulsed fraction therefore represents a compelling
evidence that the bulk of the emitted spectrum is somewhat related to
the magnetic caps, where the pulsations are formed.  The
  similarities with the pulse amplitude energy dependence of other
  AMSPs (see above), further strengthens such a conclusion.

Soft phase lags of $\sim230\mu s$ have also been observed. Soft lags
of a similar amount ($\sim 200\mu s$) were observed from the AMSP
{\saxj} by \citet{cui98}, who interpreted them in terms of Compton
downscattering of intrinsically hard photons in a $\tau\sim 10$ cloud,
which is however not observed in the X-ray spectrum. It was also
proposed that soft lags could be due to the Doppler energy shifts
introduced by the NS fast rotation \citep{Frd00,Wnb01}, or due to the
different angular distribution of the flux emitted by the blackbody
and the Comptonized emission \citep{GrlDonBar02,PouGie03}. As the lags
we observe from {\igr} are of similar amplitude than those of {\saxj}
(which spins 1.7 times faster than this source), Doppler effects are
unlikely the only physical reason behind phase lags.

\subsection{The companion star}
\label{sec:companion}

{\igr} is one of the AMSP with the longest orbital period ever
observed. From the measured mass function,
$f(m_2;m_1,i)=(m_2\sin{i})^3/(m_1+m_2)^2$ (see Table
\ref{table:spin}), it is possible to derive constraints on the mass of
the companion star, $m_2$. Here $f$,$m_1$ and $m_2$ are expressed in
solar masses. The absolute minimum on $m_2$ is reached considering an
inclination of 90$^{\circ}$, which translates in
$m_2^{(min)}\simeq0.05+0.06m_1$ for $1\leq m_1 \leq 2$. The absence of
eclipses limits the range of viable inclinations to
$i<90^{\circ}-\gamma$, where $\gamma=\arctan{(R_{2}/a)}$ is the angle
subtended by the companion star as seen from the NS, $R_2$ is the
radius of the companion star and $a$ is the binary separation. In
order for mass transfer to proceed the radius of the companion must be
comparable with the radius of the Roche Lobe, $R_2\simlt R_{L2}$,
which is estimated through the relation given by \citet{Pcz71}:
\begin{equation}R_{L2}=0.462 a [m_2/(m_1+m_2)]^{1/3}.\label{eq:pcz}\end{equation}

\begin{figure}
 \includegraphics[angle=270.0,width=\columnwidth]{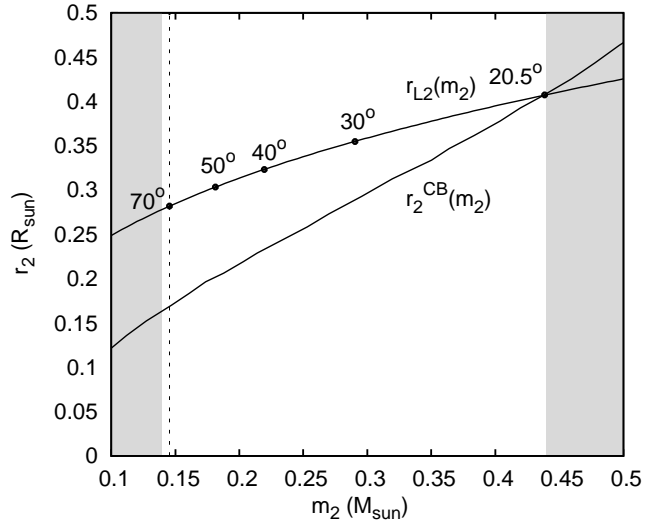}
 \caption{Size of the companion Roche Lobe (top curve) and ZAMS mass
   radius relation as from \citet[][bottom curve]{ChbBrf00}. The
   values taken by an assumed Roche Lobe filling companion star, for a
   set of inclinations and $m_1=1.4$, are also indicated. The left
   shaded region is excluded from the absence of eclipses, while the
   right shaded one from the assumption that the companion is not
   larger than its Roche Lobe. The minimum mass for $i<
   70^{\circ}$ is also plotted with a vertical dashed line. }
\label{fig:mass2}
\end{figure}

To evaluate the maximum inclination compatible with the absence of
eclipses, we calculate $\gamma$ for $R_2=R_{L2}(m_2^{(min)})$,
obtaining $i\simlt 78^{\circ}$.  The absence of dips in the X-ray
light curve further decreases the range of possible inclinations, so
that we consider $i\simlt70^{\circ}$ only, slightly increasing the
minimum companion mass to, e.g., $m_2\simgt0.15$, for $m_1=1.4$. A
main sequence companion star is clearly indicated. To see this we use
the third Kepler law in Eq.(\ref{eq:pcz}) to relate the Roche Lobe
radius to the mean density of the companion star at the observed
orbital period:
\begin{equation}
r_{L2}(m_2)=R_{L2}/R_{\odot}=0.536\;  m_2^{1/3}.
\end{equation} 
This relation is plotted in Fig.\ref{fig:mass2}, together with the
Zero Age Main Sequence (ZAMS) mass-radius relation of
\citet{ChbBrf00}, $r_2^{CB}(m_2)$. The two curves cross for
m$_2^{(max)}=0.44$ and r$_2^{(max)}=0.41$. Heavier companion stars can
be safely excluded as the Roche Lobe would be overfilled, and mass
transfer unstable. Assuming $m_1=1.4$, this solution corresponds to
i$^{(min)}=20.5^{\circ}$.

The reflection continuum and the iron K$\alpha$ line we have modelled
with a \texttt{diskline} indicate an inclination in the range
38--68$^{\circ}$ (see Table \ref{tab2} and
Sec.\ref{disc:spectrum}). This interval corresponds to
$m_2=$0.15--0.23 for m$_1=1.4$. Slightly larger values are found for
heavier NS.  The inclination estimate we have drawn from spectral
fitting thus indicates that the companion is slightly bloated with
respect to its ZAMS thermal equilibrium radius.  A possible mechanism
to drive a companion star out of thermal equilibrium is irradiation by
the compact object, but to what extent it is important is still matter
of debate \citep[see][and references therein]{Rtr08}. However,
irradiation is not strictly needed to account for the values we
observe from {\igr}, as CV-like evolution can explain it if the
companion star is slightly evolved at the onset of mass transfer (but
still before the MS turn off), or heavier than 1 M$_{\odot}$. As
\citet{PylSvn89} shown, angular momentum losses driven by magnetic
braking and gravitational radiation can still lead to a converging
system (that is, a system that evolves decreasing its orbital period),
even for slightly evolved, or heavy, companions. In these cases, the
companion radius will be systematically larger than the ZAMS
radius. To see this, we consider the evolutionary tracks calculated by
\citet{PdsRppPfh02}. While an unevolved companion with initial mass of
$m_2^{(i)}$=1 crosses the observed orbital period of {\igr} when
m$_2\simeq0.44$, as its radius is frozen to the ZAMS value, this
happens for $m_2\simeq0.16$ if it is slightly evolved, and thus
larger, at the beginning of mass transfer. Similar results are
obtained even for initially unevolved, but heavier, companion star
(the evolutionary track for a $m_2^{(i)}=1.4$ companion star passes
through the period of {\igr} when $m_2=0.15$). Values of $m_2$ in
agreement with the inclination indicated by spectral modelling can be
thus obtained, for peculiar conditions of the companion star at the
onset of mass transfer.

\subsection{The X-ray spectrum}
\label{disc:spectrum}

We presented in Sec.\ref{sec:xmmspectrum} and \ref{sec:rxtexmm} a
detailed spectral analysis of the 0.5--200 keV X-ray emission of the
AMSP {\igr}, using two different Comptonization models, and checking
the presence reflection.  The spectrum of {\igr} is modelled by four
components, (i) the accretion disc emission peaking at $\simeq 0.3$
keV; (ii) a $\simeq$ 0.6 keV blackbody of apparent radius $\simeq7$
d$_8$ km; (iii) Comptonization from a hot ($kT_e\simeq50$ keV) medium
of moderate optical depth ($\tau\simeq1.3$), of the thermal
($\simeq1.3$ keV) photons provided by a $\simeq3.5$ d$_8$ km region
(iv) Compton reflection and an iron K$\alpha$ emission line,
interpreted as coming from the accretion disc illuminated by the hard
radiation emitted from the NS surface.  The shape of the best fitting
model does not change much using different Comptonization models.  As
the model PS-refl better addresses a number of physical aspects of
this source emission (see Sec.\ref{sec:rxtexmm}) we discuss the
parameters thus obtained. Compton disc reflection is significantly
detected regardless of the particular model used to describe
Comptonization, as well as an iron line at an energy compatible with
Fe XXV-XXVI.  Even if reflection has not always been observed, such a
spectral decomposition have already proved successful in the
description of the X-ray spectrum AMSP (\citealt{GrlDonBar02}, GP05,
\citealt{Fal05,Pap09,Pat09}). In particular, the presence of disc
thermal emission has always been detected when {\xmm} high resolution
spectra extending down to $\sim$1 keV were available.  The
similarities in the X-ray emission of AMSP can be viewed as an
indication of how the physical processes that produce the bulk of
their spectra are similar, and probably related to the presence of a
magnetosphere in these systems.

\label{disc:column}

The correlation between the energy dependence of the pulsed fraction
and the spectral decomposition we have employed strongly suggests that
the Comptonizing medium surrounds the hot spots on the NS
surface. \citet{GrlDonBar02}, \citet{PouGie03}, GP05, \citet{Fal05}
and \citet{Fal05b} interpreted the similar hard components shown by
other AMSPs as coming from Comptonization in a plasma heated in the
accretion columns. The magnetic field collimates the in-falling matter
in columns of radii that are few tenths of the NS size. Even for
accretion rates $\sim 0.1$ $\dot{M}_{Edd}$ like those observed for
these sources during outbursts, the local accretion rate $\dot{m}$ can
therefore attain the Eddington level \citep{BasSun76}.  In such a
case, the radiation pressure becomes comparable to the ram pressure of
the plasma falling at supersonic velocities, and a shock may thus
form, heating the electrons up to the observed large temperatures.

The observed $\simeq 0.6$ keV blackbody component is too cold to
supply enough seed photons for the observed Comptonized spectrum.
Disentangling the temperature of seed photons from that of the
observed blackbody significantly improves the fit and ensure the
energy balance between the hot plasma and the region that provides the
seed photons. A similar result has already been obtained by GP05,
modelling the X-ray spectrum of the AMSP, {\xte}.  The size of the
region that provides the soft photons,
$R_{soft}^{\infty}=3.4^{+0.1}_{-0.4}$ d$_8$ km, is compatible with the
expected radius of an hot spot.  The radius of the observed blackbody,
$R_{BB}^{\infty}=6.3^{+0.3}_{-0.1}$ d$_8$ km, is instead of the same
order of the asymptotic value taken by the radius of the thermal burst
emission. This component is therefore emitted by a larger fraction of
the NS surface.

\label{sec:disc}
The cooler thermal component can be safely attributed to the emission
of an optically thick accretion disc. The contribution of this
component at low energy is responsible for the decrease of the pulsed
fraction. Moreover, the parameters we obtain with a \texttt{diskbb}
model nicely fit the expectations for an accretion disc around an
AMSP. Considering the range of inclinations indicated by the
reflection component (38--68$^{\circ}$), the model \texttt{diskbb}
evaluates an apparent inner disc radius $R_{in}$ in the range 18--28
d$_8$ km. Such a radius is estimated approximating the disc
temperature profile with its asymptotic behaviour $T(r)\propto
r^{-3/4}$, without accounting for the decrease of the viscous torque
at the inner boundary of the disc, and neglecting spectral
hardening. To evaluate the importance of these effects we consider the
disc model \texttt{diskpn} \citep{Grl99}. The spectral shape found by
\texttt{diskbb} is recovered after rescaling the inner disc radius
$R_{in}$ by a factor $\approx2.3$, for $m_1=1.4$, $d=8$kpc,
$i=50^{\circ}$ and $f=1.7$, where f is the ratio between the colour
and effective temperature \citep{ShmTkh95}.  The inner radius
indicated by disc emission modelling thus meets the request that an
accretion disc around an X-ray pulsar is not truncated at radii much
larger than the corotation radius, $R_{C}=(GM/4\pi^2\nu^2)^{1/3}=42.8
\: m_{1.4}$ km for {\igr}, where $m_{1.4}$ is the NS mass in units of
1.4 $M_{\odot}$. Also the ratio between the flux in the disc component
and the total flux agrees with a truncation radius a few tens of km
away from the NS star. From the virial theorem it is in fact expected
that $F_{disc}/F_{tot}\simeq R_*/2r_{in}$. As we measure
$F_{disc}/F_{tot}\simeq0.1$, the flux ratio predicts
$r_{in}\simeq5R_*$, which substantially agrees with the measured
value.

The presence of reflection features, such as an iron K$\alpha$
emission line and Compton reflection, is significantly detected in
spectral fitting.  The iron line stands at $\sim 4\sigma$ above the
continuum and has an equivalent width of $\simeq40$ eV. If modelled
with a Gaussian, its width can only be constrained to be $\simgt 0.5$
keV. As the spectrum of this source is dominated by Comptonization in
a hot and optically thin medium, it is hard to find alternatives to
disc reflection in order to explain such a width. In the inner regions
of the accretion disc the line is broadened and red-shifted by the
effects of Keplerian motion and by the gravitational influence of the
nearby NS. Modelling the feature with a \texttt{diskline} we obtain an
estimate of the inner disc radius as $r_{in}=27^{+6}_{-9}$ R$_g$,
which translates in $r_{in}=56^{+12}_{-19}$ km for a 1.4 M$_{\odot}$
NS. Even if it is rather loose due to the limited counting statistics,
also this estimate overlaps with the corotation radius, as it is
expected from accretion theories. The addition of the {\rxte} dataset
to the {\xmm} spectrum also allows the detection of the hump at $\sim
30$ keV expected from Compton reflection.  The ionised state of the
reflector is rather high, $\log\xi=3.0^{+0.4}_{-0.2}$.  It is worth to
note that the highly ionised surface in the reflector is in agreement
with the transition energy of the broad iron line at
E$_{Fe}=6.82^{+0.09}_{-0.11}$ keV, that we identify as a Ly$\alpha$
resonant transition of Fe XXV (rest-frame energy at 6.70 keV), that is
likely to be produced in a photo-ionised plasma at $\log\xi\simeq 3$
\citep{Klm04}.  Considering the values indicated by the reflection
continuum and by the iron line modelling, we give a conservative
estimate of the inclination in the range 38--68$^{\circ}$.  Also the
OVIII edge we find at a high absorption depth ($\tau=$0.1--0.3) can be
tentatively interpreted in terms of disc reflection. As its width is
$<34$ eV, the outer rings of the disc should be involved, in order to
make the rotational broadening negligible.

\subsection{The X-ray bursts}
\label{disc:burst}

In Sec. \ref{sec:burst} we have presented a spectral and temporal
analysis of the bursts observed by {\xmm}. Considering the {\it Swift}
observations, \citet{Bozzo09} concluded that the observed recurrence
times could be reconciled with the {\it persistent} flux if it is
assumed that the bursts are ignited in a pure helium environment.

This hypothesis can be checked estimating the local accretion rate
$\dot{m}$ needed to produce the fluence of the second burst, which
lags the first by $\Delta t=$40.9 ks.  One has in fact that
$\dot{m}=y/\Delta t$, where $y$ is the column depth at which the burst
is ignited. The value of $y$ can be estimated as $y=4\pi d^2
\mathcal{F}/Q_{nuc} R_*^2$ \citep[see e.g.][]{Glw08}, where
$\mathcal{F}=(1+z)\tau F_{peak}^{\infty}$ is the burst fluence, and
$Q_{nuc}$ is the energy per nucleon released during the thermonuclear
burning ($Q_{nuc}\approx 1.6$ MeV for complete burning of He into iron
group elements,\citealt{WalWoo81}). The measured values (see Table
\ref{table:burst}) yield $y=(2.6\pm0.6)\times10^8$ d$_8^2$
R$_{10}^{-2}$ g cm$^{-2}$, and $\dot{m}=(6.3\pm1.4)\times 10^3$
d$_8^2$ R$_{10}^{-2}$ g cm$^{-2}$ s$^{-1}\simeq 0.08\;R_{10}^{-1}
\dot{m}_{Edd}$, where $z=1.31$ is considered. This value of $\dot{m}$
agrees with the bolometric luminosity of the {\it persistent}
emission. Assuming that the \texttt{blackbody} and the Comptonized
component arise from the vicinity of the NS surface, and therefore the
observed fluxes have to be corrected for general relativity effects,
we in fact estimate from model PS-refl the {\it persistent} luminosity
as, $L_X^{pers}=4\pi
d^2[(1+z)^2(F_{BB}+F_{PS})+(F_{Disc}+F_{Refl})]=(1.6\pm0.1)\times
10^{37}$ d$_8^2$ erg s$^{-1}\;\simeq0.09$ m$_{1.4}^{-1}$ L$_{Edd}$.
The agreement between the {\it persistent} luminosity and the mass
accretion rate needed to explain the bursts fluence and recurrence
time, when it is assumed that all the hydrogen is steadily burnt
between bursts, suggests that bursts are triggered in nearly pure
helium environment.  It is perhaps useful to clarify that the
difference between the local accretion rate estimated by the burst
properties, $\dot{m}\simeq 0.08 \dot{m}_{Edd}$, and the rate needed
for a shock to form in the accretion column,
$\dot{m}\simeq\dot{m}_{Edd}$, arises as the first is evaluated
averaging the accretion rate over the whole NS surface, while the
latter considers only the column size. Such a treatment seems
appropriate as a magnetic field $\simlt 10^9$ G like the one expected
in an AMSP should be able to confine accreted matter in the polar caps
up to a column density of $\simeq 10^7$ g cm$^{-2}$ \citep{BrwBld98},
which is well below the value we measure. The accreted layer is
therefore able to diffuse across the whole NS surface before the burst
onset.

As no photospheric radius expansion is observed, an upper limit on the
distance can be set as $d^2\simlt{L_{PRE}}/{4\pi F_{peak}^{\infty}}$,
where the luminosity at which photospheric radius expansion is
expected, $L_{PRE}=(3.79\pm0.15)\times10^{38}$ erg s$^{-1}$, was
empirically derived by \citet{Kulk03}. Considering the peak flux of
the second burst and the associated uncertainty, we set an upper limit
on the distance to the source of $10$ kpc at a 90\% confidence
level. The radius of the thermal emission attained during the burst
decay can be identified with the NS radius, and used to put a
reasonable lower limit on the distance to the source. The measured
value $R_{app}=(7.0\pm0.3) d_8$ km can be related to the effective
radius through the relation, $R=R_{app} f_C^2(1+z)^{-1}$ \citep[see
  e.g.][]{LewvPrTam93}. The factor $f_C$ takes into account spectral
hardening, and has been estimated as $f_C\simeq1.35$ by \citet{Mdj04},
for the case of a source that does not reach the Eddington level.
Solving the previous relation for $M>1.4$ M$_{\odot}$, and considering
that a soft equation of state like the A predicts $R_*>8.5$ km and
M$\simlt 1.92$ M$_{\odot}$ for a NS spinning at the rate of {\igr}
\citep{Pan71,CokShpTeu94}, we obtain $d\simgt 6.5$ kpc, unless a
strange star is considered. For a lighter NS (M=1.57 M$_{\odot}$) this
limit increases to 6.9kpc. Given the estimated range of distances, it
is possible to safely state that {\igr} belongs to the Galactic
bulge. Also the compatibility of the measured equivalent hydrogen
absorption column with that observed from {\xte} (whose distance was
estimated to be $\simgt$ 7 kpc, \citealt{Mark02,Pap08}) supports such
a conclusion.

Pulsations are detected throughout the burst decay at a similar
amplitude than that seen during the {\it persistent} emission, while
during the first few seconds they disappear, probably because of the
telemetry limitations of the EPIC-pn.  Their phase is also compatible
within 0.1--0.2 cycles with that of pre-burst pulsations, suggesting
that magnetic channelling of accreted matter and the burst onset
happen at a similar location on the NS surface. This indication is
strengthened by the analysis performed by Riggio et al. (2010,
submitted) over RXTE data. They find in fact that, with the exception
of the first two seconds since the burst onset, the burst oscillations
are phase locked to pre-burst pulsations within 0.1 cycle.  Phase
locking between burst and non-burst oscillations of an AMSP has
already been observed by \citet{Wts08}, at a much larger accuracy
($\simeq$0.01 cycles) than that made available by the data presented
here.  As they discuss (see also references therein), the most
appealing interpretation is in terms of the effect a temperature
gradient between the fuel impact point on the surface, and the rest of
the surface, may have on the ignition conditions, or on its
development. A temperature gradient during the {\it persistent}
emission is indeed indicated by our spectral modelling, which requires
the region feeding the accretion columns of soft photons to be hotter
than the rest of the surface. A firm explanation of the correlation
between pulsations observed during bursts and during the {\it
  persistent} emission is anyway still missing.

\section*{Acknowledgments}

We thank N.Schartel, who made possible this ToO observation in the
Director Discretionary Time, and the {\xmm} team who performed and
supported this observation. We also thank M.Falanga for useful
discussions. This work is supported by the Italian Space Agency,
ASI-INAF I/088/06/0 contract for High Energy Astrophysics, as well as
by the operating program of Regione Sardegna (European Social Fund
2007-2013), L.R.7/2007, ``Promotion of scientific research and
technological innovation in Sardinia''.

\bibliographystyle{mn2e}
\bibliography{biblio}

\end{document}